\documentclass[prd,amsmath,amssymb,showpacs,showkeys,floatfix,
               nofootinbib,eqsecnum,
               preprint,12pt,tightenlines,fleqn
               ]{revtex4-1}
\usepackage{graphicx}

\def\al{\alpha}

\def\ga{\gamma}
\def\de{\delta}

\def\ze{\zeta}

\def\ka{\kappa}

\def\vp{\varphi}

\def\De{\Delta}

\newcommand{\beq}{\begin{equation}}
\newcommand{\eeq}{\end{equation}}
\newcommand{\bea}{\begin{eqnarray}}
\newcommand{\eea}{\end{eqnarray}}
\newcommand{\bse}{\begin{subequations}}
\newcommand{\ese}{\end{subequations}}

\newcommand{\BB}{\big}
\newcommand{\BBB}{\Big}

\newcommand{\nn}{\nonumber}
\def\vev#1{\langle {#1}\rangle}
\def\ket#1{|{#1}\rangle}
\def\bra#1{\langle{#1}|}

\usepackage{amsmath}
\usepackage{amsfonts}
\usepackage{amssymb}

\newcommand{\bM}{\begin{pmatrix}}
\newcommand{\eM}{\end{pmatrix}}

\newcommand{\Ha}{\frac{1}{2}}

\def\ol#1{\overline{#1}}

\def\CvB{CNB}
\def\vE{\mathbf{v}_{\oplus}}    
\def\vv{\mathbf{v}}             
\def\qv{\mathbf{q}}
\def\vvh{\widehat{\mathbf{v}}}      
\def\vu{\mathbf{u}}       %
\def\vq{\mathbf{q}}       %
\def\vp{\mathbf{p}}       %
\def\R{\text{rel}}
\def\NR{\text{non-rel}}

\begin{document}

\noindent  Phys. Rev. D 93, 053004 (2016)
\hfill  arXiv:1512.00817\newline\vspace*{4mm}

%

\title{Neutrino refraction by the cosmic neutrino background}
\author{J.S. D\'iaz}
\email{jorge.diaz@kit.edu}
\author{F.R. Klinkhamer}
\email{frans.klinkhamer@kit.edu}
\affiliation{Institute for
Theoretical Physics, Karlsruhe Institute of
Technology (KIT), 76128 Karlsruhe, Germany\\}

\begin{abstract}
\noindent
\vspace*{-4mm}\newline
We have determined the dispersion relation of a neutrino test particle propagating in the cosmic neutrino background.
Describing the relic neutrinos and antineutrinos from the hot big bang as a dense medium,
a matter potential or refractive index is obtained.
The vacuum neutrino mixing angles are unchanged, but
the energy of each mass state is modified.
Using a matrix in the space of neutrino species,
the induced potential is decomposed into a part which produces signatures in beta-decay experiments and another part which modifies neutrino oscillations.
The low temperature of the relic neutrinos makes a direct detection  extremely challenging. From a different point of view,
the identified refractive effects of the cosmic neutrino background constitute
an ultralow background for future experimental studies of nonvanishing
Lorentz violation in the neutrino sector.

\end{abstract}

\maketitle
\section{Introduction}

The hot big bang model~\cite{Gamow:1946,Gamow:1948,AlpherHerman:1949}
predicts the existence of relic photons in the Universe,
remnants of the primordial plasma which became transparent about 380,000 years after the inferred big bang singularity.
Today, these photons appear as the cosmic microwave background radiation
(CMBR)~\cite{PenziasWilson1965,Mather-etal1990},
the properties of which 
have been measured with great
precision~\cite{CMBR.COBE:1992,CMBR.WMAP:2013,CMBR.Planck:2015}.
The hot big bang model also predicts that
neutrinos and antineutrinos  of all flavors become free-streaming particles
at approximately one second after the big bang.
This cosmic neutrino background (\CvB)  remains undetected for now.
The direct detection of the CNB  constitutes a challenging task for both particle physics and cosmology.

The study of the CMBR has allowed us to reach a remarkable understanding of the early stages of the Universe. Similarly,
the study of the \CvB\ would provide unprecedented insight into the
1-sec-old   
Universe, that is, the Universe from
before the nucleosynthesis of the primordial elements.
The potential importance of the \CvB\ has led to several detection proposals; some ideas are currently being implemented,
others remain out of experimental
reach~\cite{Langacker:1983,Gelmini:2004,Yanagisawa:2014}.

In this work,
the possible effects of the \CvB\ on the propagation of
neutrino test particles are studied.
We consider the \CvB\ as a medium in which neutrino test particles propagate and we determine the effective potential which modifies their dispersion relation.
This paper is organized as follows.
The thermal history of cosmic neutrinos is summarized in Sec.~\ref{Sec.CvB},
while techniques for direct detection of the \CvB\ are reviewed
in Sec.~\ref{Sec.DirectDet}.
The refractive effects of the \CvB\ are calculated in Sec.~\ref{Sec.CvBeffects},
and the corresponding experimental signatures are discussed
in Sec.~\ref{Sec.Exp}. Concluding remarks are presented in
Sec.~\ref{Sec.summary}. Throughout,
we use natural units $c=\hbar=k_B=1$, unless otherwise stated.

\section{cosmic neutrino background}
\label{Sec.CvB}

The Universe was a hot plasma at around a fraction of a second after
the big bang~\cite{Weinberg:1972}.
There was thermal equilibrium between electrons, positrons,
and photons, primarily due to their electromagnetic interactions.
Left-handed neutrinos and right-handed antineutrinos
shared this thermal equilibrium due to
scattering processes mediated by weak interactions.
When the temperature decreased to a few MeV,
the weak-interaction scattering rate dropped
below the Hubble expansion rate.
As a result, the neutrinos decoupled from the plasma, henceforth
evolving with an independent ``temperature''
$T_\nu$.  This gas of free-streaming neutrinos constitutes the \CvB.

After the neutrino decoupling,
the photon temperature continued to decrease and soon the energy of the photon gas dropped below the threshold for electron-positron
production ($\gamma+\gamma$  $\to$  $e^{-}+e^{+}$).
The remaining pairs of charged leptons
annihilated ($e^{-}+e^{+}$ $\to$ $\gamma+\gamma$)
and transferred their entropy to the photon gas.
This extra entropy made the photons cool less than the neutrinos
by a factor $(11/4)^{1/3}$.
Since the neutrinos were relativistic at the time of decoupling,
they preserve, to a good approximation, the form of the
Fermi--Dirac distribution function $f_{\nu_i}(\vq,T_\nu)$.
From this function $f_{\nu_i}$, the number density of a mass eigenstate $\nu_i$ can be determined as
\beq\label{n_v(T)}
n_{\nu_i}(T_\nu) = \int \frac{d^3q}{(2\pi)^3} \; f_{\nu_i}(\vq,T_\nu)
= \frac{3\ze(3)}{4\pi^2}\; T_\nu^3.
\eeq
Since the CMBR today has a blackbody
spectrum~\cite{Mather-etal1990} with $T^0_\text{CMBR}\simeq0.24$ meV,
we find that the present \CvB\ temperature is
\beq\label{T-0-CNB-numerical}
 T^0_\text{\CvB}=(4/11)^{1/3}\;T^0_\text{CMBR}\simeq 0.17 \,\text{meV}.
\eeq
The temperature value \eqref{T-0-CNB-numerical}
implies that the present neutrino density \eqref{n_v(T)} is 56 cm$^{-3}$ per neutrino state,
with a total of 336 cm$^{-3}$ for three flavors of
left-handed neutrinos and right-handed antineutrinos.

This rather high density makes neutrinos relevant for the evolution of the Universe.
In fact,
each neutrino species contributes to the total energy density and leaves measurable imprints on the properties of the
CMBR~\cite{CMBR.WMAP:2013,CMBR.Planck:2015,Follin:2015}.
These signatures have allowed the indirect detection of the \CvB.
For large enough neutrino masses compared to
the value \eqref{T-0-CNB-numerical}, these background neutrinos would
constitute a highly nonrelativistic gas due to their low temperature,
which would make a direct detection today a challenging task.
For such a nonrelativistic state of the \CvB,
there may be observable differences depending
on the Dirac or Majorana nature of the neutrino mass.

\section{Direct detection of the \CvB}
\label{Sec.DirectDet}

One of the most promising methods under consideration today for the direct detection of the \CvB\ is neutrino capture~\cite{Weinberg:1962}.
A nonrelativistic electron neutrino from the \CvB\   
can be captured by a beta-decaying nucleus, $\nu_e+N\to N'+e^-$.
This reaction has no threshold energy for the incoming neutrino and its experimental signature corresponds to a narrow peak in the beta spectrum centered at an energy $2m_\nu$ above the
end point  
of the spectrum,
where $m_\nu$ is the effective mass of the captured neutrino.
This detection method is mainly limited by the energy resolution of the electron spectrum and by the mass of the target nuclei.
Given the relevant observable for this type of search,
experiments designed for direct measurements of the neutrino mass appear as potential
instruments
for \CvB\ detection. However,
the current specifications for experiments such as the Karlsruhe Tritium Neutrino experiment (KATRIN)~\cite{KATRIN}
and the Microcalorimeter Arrays for a Rhenium Experiment (MARE)~\cite{MARE}
indicate that the capture rate would be too low for a successful detection of the \CvB~\cite{NCB1,NCB2,NCB3}.
A dedicated experiment under development,
which intends to implement modern  technologies for high resolution as well as a massive tritium target, is
the Princeton Tritium Observatory for Light, Early-Universe, Massive-Neutrino Yield (PTOLEMY)~\cite{PTOLEMY}.

The capture rate can be enhanced by gravitational
clustering~\cite{Gershtein:1966,Klinkhamer:1981,Bond:1980,Bond:1983,Ringwald:2004} and increased by a factor
of 2  
if neutrinos are Majorana particles
instead of Dirac particles~\cite{Long:2014}.
The use of a polarized target has also been suggested because it could enhance some of the \CvB\ detection signals~\cite{Lisanti:2014}
and gravitational focusing has been shown to provide key experimental signatures~\cite{Safdi:2014}.

A different approach which could directly reveal the existence of the \CvB\
is based on the coherent elastic scattering of \CvB\ neutrinos with a target.
Depending on the spin state of an electron in a terrestrial laboratory,
an energy difference is produced by the motion of the laboratory with respect to \CvB~\cite{Stodolsky:1975}.
This proposed detection method is the only one
which is linear in the weak coupling $G_F$~\cite{Langacker:1983}.
The method requires, however, a significant difference between the number density of neutrinos and that of antineutrinos
(i.e., a significant density of lepton number $L$ in the CNB).

Other detection methods suggest
using  
the \CvB\ as a target for high-energy astrophysical neutrinos which could annihilate with antineutrinos in the \CvB\ through the $Z$ resonance and
produce a dip in the high-energy neutrino flux~\cite{Zburst}.

\section{\CvB\ effects on neutrino propagation}
\label{Sec.CvBeffects}

The effects of a dense medium on neutrino propagation were first studied
by Wolfenstein~\cite{Wolfenstein:1978}.
From this work, it is known that the neutrino dispersion relation gets
modified by coherent  forward scattering.
In this section, we calculate how a neutrino test particle is affected by the presence of
the cosmic background of neutrinos and antineutrinos.
In particular,
we will determine the observable \CvB\ signals in neutrino experiments.

\begin{figure}
\centering
\includegraphics[width=0.75\textwidth]{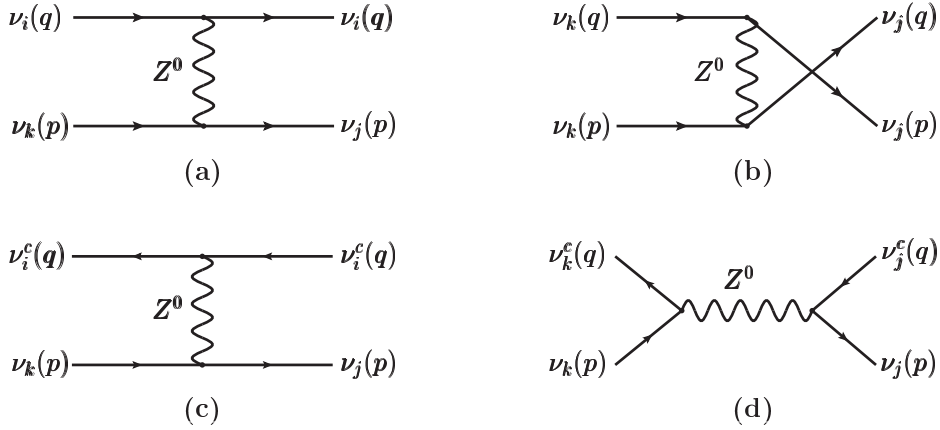}
\caption{Feynman diagrams for the elastic scattering of a neutrino test particle
of momentum $p$ with background neutrinos and antineutrinos
of momentum $q$.
}
\label{Fig:Diagrams}
\end{figure}

The neutrino test particle is taken as a relativistic state which can interact with neutrinos in a medium via neutral currents.
For neutrinos with energies below the mass of the $Z$ boson,
the Hamiltonian density describing this neutrino-neutrino interaction can be written as a four-fermion interaction,
\beq\label{Hvv}
\mathcal{H} = \frac{G_F}{2\sqrt2}
\sum_{l,m}
\BBB[\ol\nu_l\ga^\mu(1-\ga_5)\nu_l\BBB]\,
\BBB[\ol\nu_m\ga_\mu(1-\ga_5)\nu_m\BBB],
\eeq
where the indices $l,m \in \{1,2,3\}$ label the field components in the mass basis. This Hamiltonian density has been studied in detail
for, e.g., neutrino interactions in the early universe and
collective behavior of neutrinos in the dense medium of supernova
explosions~\cite{NotzoldRaffelt,Pantaleone:1992a,Pantaleone:1992b,%
Enqvist:1990,Kostelecky:1993,KosteleckySamuel,Kuo:1989,Volpe:2013}.
The effective Hamiltonian density \eqref{Hvv}
leads to exact results for the tree-level diagrams (a) and (c) in Fig.~\ref{Fig:Diagrams} because the
forward scattering makes the momentum carried by the $Z$ boson vanish.
We consider the energy of all particles to be small
compared to the $Z$-boson mass and
the approximation \eqref{Hvv} is valid only at leading order for
the tree-level diagrams (b) and (d)  in Fig.~\ref{Fig:Diagrams}.

For the medium description of the background neutrinos in
Fig.~\ref{Fig:Diagrams}a,
we sum the Hamiltonian density \eqref{Hvv} over neutrino states $i$ and
momentum $\vq$ and average over helicity $h$,  giving
\beq\label{<Hvv>a}
\ol{\mathcal{H}}_\text{(a)} =
\Ha \;\sum_{h=\pm 1/2}\; \sum_{i=1}^{3}\;\int d^3q   
\,f_{\nu_i}(\vq,T_\nu)\;\bra{\nu_i(\vq,h)}\mathcal{H}\ket{\nu_i(\vq,h)},
\eeq
where $f_{\nu_i}(\vq,T_\nu)$ is the Fermi--Dirac distribution function from
Sec.~\ref{Sec.CvB}. Since we will consider the possibility
of a nonrelativistic background of massive
neutrinos,   
we avoid the use of the terms left-
and right-handed states because helicity and chirality are equivalent only in the relativistic regime.

For a neutrino test particle of momentum $\vp$ and
negative helicity $h=-1/2$ propagating
through the CNB, the experienced potential is obtained by
integrating the averaged Hamiltonian density over space,
\beq
V^{(\text{a})}_{jk}
= \int d^3x \, \bra{\nu_{j}(\vp,-1/2)}\ol{\mathcal{H}}_{(\text{a})}(x) \ket{\nu_{k}(\vp,-1/2)}.
\eeq
Following the same procedure for the other diagrams in
Fig.~\ref{Fig:Diagrams},
we find that the total neutrino potential,
\beq\label{Vabcd}
V_{jk}=V^{(\text{a})}_{jk} +V^{(\text{b})}_{jk} +V^{(\text{c})}_{jk} +V^{(\text{d})}_{jk}\,,
\eeq
is diagonal in the space of neutrino species:
\beq\label{V-diagonal}
V_{jk}=V_{jj}\;\de_{jk}\,,
\eeq
without sum over $j$ on the right-hand side.
This diagonality implies that the potential modifies the neutrino dispersion relations while leaving the mixing unchanged. Explicitly,
the diagonal $jj$ entry can be written as
\begin{widetext}
\bea\label{VjjD(total)}
V_{jj} &=& \frac{G_F}{\sqrt2} \sum_{i} (1+\de_{ij})
\BBB[
\BB(1 - \widehat{\vp}\cdot\vev{\vv_i}\BB)
\BB(n^{-}_{\nu_i} + n^{+}_{\nu_i} - n^{-}_{\nu_i^c} - n^{+}_{\nu_i^c}\BB)
\nn\\
&&\hspace{2.7cm}
+ \,
\BB(\vev{|\vv_i|} - \widehat{\vp}\cdot\vev{\vvh_i} \BB)
\BB(n^{-}_{\nu_i} - n^{+}_{\nu_i} + n^{-}_{\nu_i^c} - n^{+}_{\nu_i^c}\BB)
\BBB]
,
\eea
\end{widetext}
where $\widehat{\vp}$ is the direction of propagation of the neutrino
test particle, $\vv_i \equiv \qv_i/E_i$  is
the velocity of the background neutrino particle of type $i$,
and the factors $ n^{\pm}_{\nu_i,\nu_i^c}$ denote the number density of the background neutrinos ($\nu_i$) or antineutrinos ($\nu_i^c$) of helicity $h=\pm1/2$.

If the background is isotropically distributed then $\vev{\vv_i}=\vev{\vvh_i}=0$.
This isotropic distribution is expected to hold in the \CvB\ rest frame.
For an Earth-based laboratory, the velocity distribution becomes anisotropic due to the neutrino wind in the laboratory frame.
Denote the velocity of the background neutrinos in the \CvB\ frame
by $\vu_i$ and define $u_i\equiv|\vu_i|$. Then,
the normalized and isotropic velocity distribution $f^{\text{\CvB}}_{\nu_i}(\vu_i)=f^{\text{\CvB}}_{\nu_i}(u_i)$
can be related to the distribution in a terrestrial frame by $f^{\text{\CvB}}_{\nu_i}(u_i) = f^{\oplus}_{\nu_i}(\vv_i)$.
The Earth moves at a nonrelativistic speed $\vE$ relative to the CMBR frame.
Assuming that the same applies to the \CvB\ rest frame,
the relation between the velocities $\vu_i$ and $\vv_i$ is given
by the Galilean transformation $\vu_i = \vv_i + \vE$.
The average velocity in the laboratory frame becomes
\bea
\vev{\vv_i} &=& \int d^3v_i \, f^{\oplus}_{\nu_i}(\vv_i) \, \vv_i
= \int d^3u_i \, f^{\text{\CvB}}_{\nu_i}(u_i) \, (\vu_i - \vE)
\nn\\
&=& - \vE,
\eea
which shows that the effective neutrino wind is a consequence of the motion of the laboratory relative to the \CvB\ rest frame.
The same argument applies to the antineutrinos of the CNB.

If neutrino masses have a value below the present \CvB\ temperature, $m_{\nu_i}\ll T^0_{\text{\CvB}}$,
then the background particles can be described as a relativistic gas of neutrinos and antineutrinos.
A relativistic neutrino test particle with negative helicity moving through the CNB experiences the following potential from \eqref{VjjD(total)}:
\bea
V_{jj}^{\;(\R)} &=& \sqrt2\,G_F \BB(1 + \widehat{\vp}\cdot\vE \BB)
\sum_{i} (1+\de_{ij})
\BB(n^{-}_{\nu_i} - n^{+}_{\nu_i^c} \BB),
\label{VjjD(total)_R}
\eea
where only negative-helicity neutrinos and positive-helicity antineutrinos
of the CNB contribute, as expected in this limit
(also taking the theory to be Lorentz and CPT invariant).
Similar to the Stodolsky effect~\cite{Stodolsky:1975},
the potential vanishes in the absence of a neutrino asymmetry.
Additionally,
this potential is proportional to the Fermi constant and the density of the background,
which indicates that the \CvB\ simply introduces a neutrino matter
effect~\cite{Wolfenstein:1978}.
This phenomenon can also be interpreted as the existence of a nondispersive index of refraction due to the \CvB.

The case of a nonrelativistic background, $T_\nu\ll m_{\nu_i}$, requires some care in the determination of the average velocities entering the potential \eqref{VjjD(total)}. In this limit,
neutrinos and antineutrinos have a Maxwell-Boltzmann velocity distribution,
which gives $\vev{|\vv_i|}_\NR = \sqrt{8T_{\nu}/(\pi m_{\nu_i}) }$.
Hence,
the term $\vev{\vvh_i}_\NR$ takes the form $\vev{\vvh_i}_\NR=-\ka\,\vE$,
with a numerical factor $\ka=\mathcal{O}(10^2)$ for $m_{\nu_i}\approx 1$ eV.
A similar enhancement factor $\ka$ appears
in modern versions of the Stodolsky effect~\cite{Duda:2001}.
In addition, for nonrelativistic background particles which become gravitationally clustered,
their helicities become totally mixed and the potential becomes half
that of the relativistic case, $V^{(\NR)} = V^{(\R)}/2$.

Remark that the neutrino wind produces an effect equivalent
to that of a background field coupled to the neutrino test particle.
The matter effects widely studied for neutrinos propagating in the Sun and through the Earth's crust produce an isotropic index of refraction.
In contrast, the neutrino wind of the \CvB\ in the laboratory frame makes the corresponding background field to have both space and time components.
In this sense, the \CvB\ mimics the behavior of a background field which breaks Lorentz invariance.
Direct calculation of the corresponding effect of the \CvB\ on an antineutrino test particle shows that the potential differs by an overall sign. In fact, dimensional analysis shows that the potential
is equivalent to the controlling coefficient of a
dimension-3  
operator for CPT-odd Lorentz violation at the Lagrangian level~\cite{KM:2012}.
Precisely this kind of CPT-odd effect has been shown to arise in
condensed-matter-type models with Fermi-point splitting~\cite{Klinkhamer:2004hg,Klinkhamer:2004hm,Klinkhamer:2006yi}.

\section{Experimental prospects}
\label{Sec.Exp}

In this section, we explore the potential signatures of the \CvB\ as a background medium which affects the kinematics of neutrinos in experiments.
The potential \eqref{V-diagonal} with \eqref{VjjD(total)} is
independent of the momentum of the neutrino test  particle in the CNB
rest frame, which implies that the neutrino group velocity is unaffected.
Therefore, these CNB effects are unobservable in time-of-flight studies,
at least as long as the Earth velocity with respect to the CNB
rest frame can be neglected.
As mentioned above,
one of the important features of the potential \eqref{V-diagonal} is the structure in the space of neutrino species.
Its diagonal form implies that the matrix which mixes the three neutrino species is unaffected by the \CvB.
But the energy of each of the species will differ from the vacuum case.

The full potential can then be decomposed into two relevant parts,
$V=V_1+V_2$. The first part $V_1$ is a flavor-blind component which changes the three neutrino species equally and
the second part $V_2$ modifies the species separately.
As an illustration,
consider the explicit matrix form of the relativistic potential \eqref{VjjD(total)_R},
which can be written as the sum of the following two components:
\bse\label{VjjD(total)_R_matrix}
\bea\label{VjjD(total)_R_matrix-1}
V^{\;(\R)}_1 &=& \sqrt2\,G_F \BB(1 + \widehat{\vp}\cdot\vE \BB)\,
\left(\sum_{i=1}^{3} \De n_{\nu_i}\right)\,
\bM \;\;1\;\;&\;\;0\;\;&\;\;0\;\; \\
0&1&0 \\ 0&0&1 \eM,
\\[2mm]\label{VjjD(total)_R_matrix-2}
V^{\;(\R)}_2 &=& \sqrt2\,G_F \BB(1 + \widehat{\vp}\cdot\vE \BB)
\bM \De n_{\nu_1} &0&0 \\ 0& \De n_{\nu_2} &0 \\ 0&0& \De n_{\nu_3} \eM
,
\eea
\ese
with
\beq
\De n_{\nu_i} \equiv n^{-}_{\nu_i} - n^{+}_{\nu_i^c}.
\eeq
The first term \eqref{VjjD(total)_R_matrix-1}
is proportional to the identity matrix
and produces no effects on neutrino oscillations.
This flavor-blind effect of the \CvB\
could, however, introduce modifications in processes involving the neutrino phase space such as weak decays.
The second term \eqref{VjjD(total)_R_matrix-2} modifies the energy of each mass state differently and affects the phase of the transition probability in neutrino oscillations.

To   
quantify the relevant experimental signatures of the potential \eqref{VjjD(total)} in
neutrino experiments we can take advantage of the existing studies of Lorentz violation in the neutrino sector
by making the appropriate identification between our potential \eqref{VjjD(total)} and the coefficients
for CPT-odd Lorentz violation.
The neutrino wind determines a preferred direction in space,
breaking invariance under rotations.
For a neutrino test particle produced and detected in a terrestrial experiment
the term $\widehat{\vp}\cdot\vE $ introduces a time dependence because $\widehat{\vp}$
is fixed in the laboratory frame while the orientation of $\vE$ oscillates
due to Earth's rotation with sidereal frequency
$\omega_{\oplus}\approx 2\pi/(\text{23 h 56 m})$.
Such sidereal variations may lead to key experimental signatures.

The flavor-blind part of the potential can be identified with the so-called oscillation-free Lorentz-violating coefficients
$(a_{\text{of}}^{(3)})^\al$ which could produce observable effects in
beta-decay   
experiments~\cite{Diaz:2013a}.
The experimental signature corresponds to a distortion of the full beta spectrum in the form of an extra
contribution to the event rate with a different energy dependence.
Experiments of neutron decay~\cite{Diaz:2014d}
and the two-neutrino mode of double beta decay~\cite{Diaz:2014b}
may look for this type of spectral distortion.
Recently, a study of this type
was  
performed by the EXO-200 Collaboration \cite{EXO:2016}.
The anisotropies produced by the motion of the laboratory relative to the rest frame of the \CvB\ would introduce
unconventional modifications to asymmetries used for the study of angular correlations in neutron decay and
changes in the spectrum
end point  
of experiments for direct neutrino-mass measurements such as KATRIN~\cite{Diaz:2014d}.
Since the CNB effects on beta decays are solely a consequence of the modified phase space of the emitted antineutrino,
the sensitivity to the effects of the \CvB\ is expected to be limited.

The second part of the potential leads to modified neutrino oscillations.
As a natural interferometer,
the phenomenon of neutrino oscillations offers better sensitivity than weak decays. Once again,
we can borrow results for CPT-odd Lorentz violation in neutrino oscillations,
which allow us to directly identify the components of the potential \eqref{VjjD(total)} with the
different coefficients for CPT-odd Lorentz violation
denoted $(a_L)^\al$~\cite{Diaz:2009,Diaz:2015b}.
Recall that the potential \eqref{VjjD(total)} is expressed in the mass basis
and that the matrix in the flavor basis follows by transforming with the conventional mixing matrix.
This transformed potential gives rise to a new characteristic
neutrino-oscillation length scale of order
\beq
\label{L-osc}
L_{\text{osc}}^\text{\;(\R.\;CNB)}\sim (\sqrt2\,G_F \De n_{\nu_i})^{-1}.
\eeq

Regarding the numerical value of the potential  \eqref{VjjD(total)_R_matrix},
the magnitude of the neutrino asymmetries $|\De n_{\nu_i}|$
can be expected to be similar to the magnitude of the baryon asymmetry
$|n_B|= n_\ga\,\mathcal{O}(10^{-10})$.
(Efficient sphaleron-type processes~\cite{tH1976PRL,KM1984}
for the very high temperatures of
the very early universe
give $B+L=0$ at later temperatures $T\ll M_W$,  
and hence $L=-B$  in the present universe.)
But maybe neutrinos can avoid this estimate as they are electrically neutral particles and may have unusual interactions in the early
universe (perhaps with so-called ``sterile'' neutrinos).
A large neutrino asymmetry would give rise to extra relativistic species contributing to the expansion rate of the Universe.
This allows for experimental bounds on the degeneracy parameters
$\xi_{i}$ of all flavors using cosmological data.
Tight constraints on the electron-neutrino component can be obtained by searching for modifications to the freeze-out neutron-to-proton ratio,
as these modifications modify the abundance of light elements.
The resulting constraints are $-0.01<\xi_e<0.22$, $|\xi_{\mu,\tau}|<2.6$
(at 95\% C.L.)~\cite{Hansen:2001,Steigman:2007}.
For $|\xi_i|\sim0.01$, the neutrino-antineutrino density difference becomes
\beq
\label{Delta-nu-i-numerical}
|\De n_{\nu_i}| = \frac{T^3_{\nu}}{6\pi^2}  \;|\xi_i| \; \BB(\pi^2 + \xi^2_i\BB) \sim 10^{-41} \text{ GeV}^3.
\eeq
With \eqref{Delta-nu-i-numerical}, the order of magnitude of
the effective potential \eqref{VjjD(total)_R_matrix},
with the superscript `CNB' added, would be
\beq\label{V-magnitude}
\left| \det\,V^\text{(\R.\;CNB)} \right|^{1/3}\sim 10^{-46} \text{ GeV},
\eeq
or even a factor $10^7$ smaller if the
degeneracy parameter $|\xi_i|$ is of order $10^{-9}$.
To date,
the highest sensitivity to CPT-odd Lorentz violation in
the neutrino sector is at the level of $10^{-24}$ GeV
(at 95\% C.L.)   
from
atmospheric neutrino oscillations~\cite{LV_IceCube,LV_SK}.
Hence, the signatures of the \CvB\ with a typical strength of at most
\eqref{V-magnitude}
lie many orders of magnitude beyond the current sensitivity.
Incidentally, the oscillation length \eqref{L-osc}
has a numerical value from \eqref{Delta-nu-i-numerical} of the
order of $10^{30}\,\text{m}$,
which also makes clear how small the CNB effects are.

Despite the fact that the expected CNB signal seems too small for a direct detection in the near future,
it is important to consider the refraction effect produced by the \CvB\ as a possible background for future searches of CPT violation in the neutrino sector.
Furthermore, unexpected effects could enhance the neutrino density.
For instance, a neutrino overdensity may arise from gravitational
clustering~\cite{Gershtein:1966,Klinkhamer:1981,Bond:1980,Bond:1983,Ringwald:2004}
or
a nonthermal neutrino background component \cite{Chen:2015}.
Moreover,
the study presented here only considers the leading-order description
of the weak interactions.
A more detailed study including other terms in the $Z$-boson propagator would introduce terms which grow with the energy of the neutrino
test particle and are proportional to the energy density of the \CvB\ rather than the number density asymmetry~\cite{NotzoldRaffelt,Lunardini:2000}.
These terms would mimic the effects of the controlling coefficient of a dimension-4   
operator for CPT-even Lorentz violation~\cite{KM:2012}.

\section{Summary}
\label{Sec.summary}

In this article, we have studied the effects of the \CvB\ on the propagation of a neutrino test particle. In particular,
we considered terrestrial neutrino experiments to identify the potentially observable key signatures of the cosmic relic neutrinos and antineutrinos.

Using a thermal-background formalism for neutrino-neutrino interactions we have determined the effective potential of a neutrino test particle.
This effective potential due to the CNB
modifies the dispersion relation of the
neutrino test particle.
With regard to the neutrino-species structure,
the effective potential splits into a flavor-blind component which could produce experimental signals in beta-decay experiments
and a component which would alter the transition probability in neutrino oscillations.

A direct correspondence can be established between the effective potential due to the interaction with the \CvB\ and the parameters of CPT-odd Lorentz violation in the neutrino sector.
Current limits from these studies show that a direct detection of the \CvB\
potential \eqref{V-diagonal} with \eqref{VjjD(total)}
is extremely challenging
(meaning unfeasible at present).  
Still, the effects described could play a role in future
experimental studies of unconventional physics with neutrinos. Alternatively, similar effects could occur for other, hypothetical types of relic particles.

\section*{\hspace*{-5mm}ACKNOWLEDGMENTS}
\noindent
This work was supported in part by
the German Research Foundation (DFG)   
under Grant No. KL 1103/4-1.

\end{document}